\documentclass[twocolumn,twoside,slac_two]{revtex4}
\usepackage{graphicx}
\usepackage{fancyhdr}
\usepackage{graphicx}
\usepackage[a4paper,breaklinks,dvipdfm]{hyperref}

\newcommand{\fermi}{\emph{Fermi}~}

\begin{document}

\title{Gamma-ray band and multi-waveband variability of blazars with the Fermi Large Area Space Telescope}


\author{Stefano Ciprini}
\affiliation{1. ASI Science Data Center, Frascati, Roma, Italy}%
\affiliation{2. INAF Observatory of Rome, Monte Porzio Catone, Roma, Italy}
\affiliation{\emph{(on behalf of the Fermi LAT collaboration)}.}

\begin{abstract}
The Fermi Gamma-ray Space Telescope, as an all-sky survey and monitoring mission, is producing daily/weekly sampled gamma-ray light curves for dozens of blazars and other high-energy sources. Highlights on MeV-GeV gamma-ray variability properties of these sources are reported together with a few remarks about some multi-waveband observing campaigns led by Fermi and targeted to known or newly discovered gamma-ray blazars.
\end{abstract}

\maketitle

\thispagestyle{fancy}


%
\section{Gamma-ray MeV-GeV band variability of \fermi blazars}   
%
%

%
\begin{figure}[b!!!]
\vspace*{-0.4 cm}
\begin{center}
\hskip -0.3cm 
\resizebox{6.7cm}{!}{\rotatebox[]{0}{\includegraphics{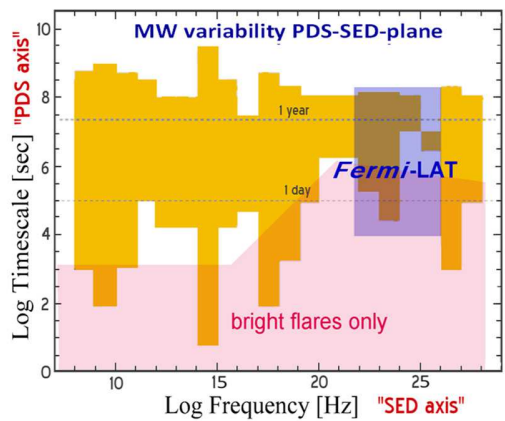}}}
\vspace*{-0.4 cm}
\caption{Multi-waveband (a.k.a. multiwavelength, MW, multifrequency) variability observed from radio-band to TeV $\gamma$-rays. Yellow shaded polygon roughly represents the timescales range probed for variability at different energy bands (i.e., measures in the power density spectra versus the spectral energy distribution space, PDS-SED plane). The \fermi LAT instrument is exploring a large portion of this plane. Adapted from \citet{wagner01}.}
  \label{fig:SED-PDS_plane}
\end{center}
\vspace*{-0.4 cm}
\end{figure}
%

The Large Area Telescope (LAT), on board the \fermi Gamma-ray
Space Telescope \cite{atwood09}, is a pair-conversion $\gamma$-ray
telescope, sensitive to photon energies from about 20 MeV up to $>300$
GeV. The LAT consists of a tracker (two sections, front and back), a calorimeter and an
anti-coincidence system to reject the charged-particle
background. \fermi LAT, working in all-sky survey mode, is an optimal hunter of high-energy flares, transients and new gamma-ray sources, and is an unprecedented monitor of the variable $\gamma$-ray sky, thanks to the large peak effective area, wide field of view ($\approx 2.4$~sr), improved angular resolution and sensitivity.

%
\begin{figure}[b!!!]
\vspace*{-0.6 cm}
\begin{center}
\resizebox{8.6cm}{!}{\rotatebox[]{0}{\includegraphics{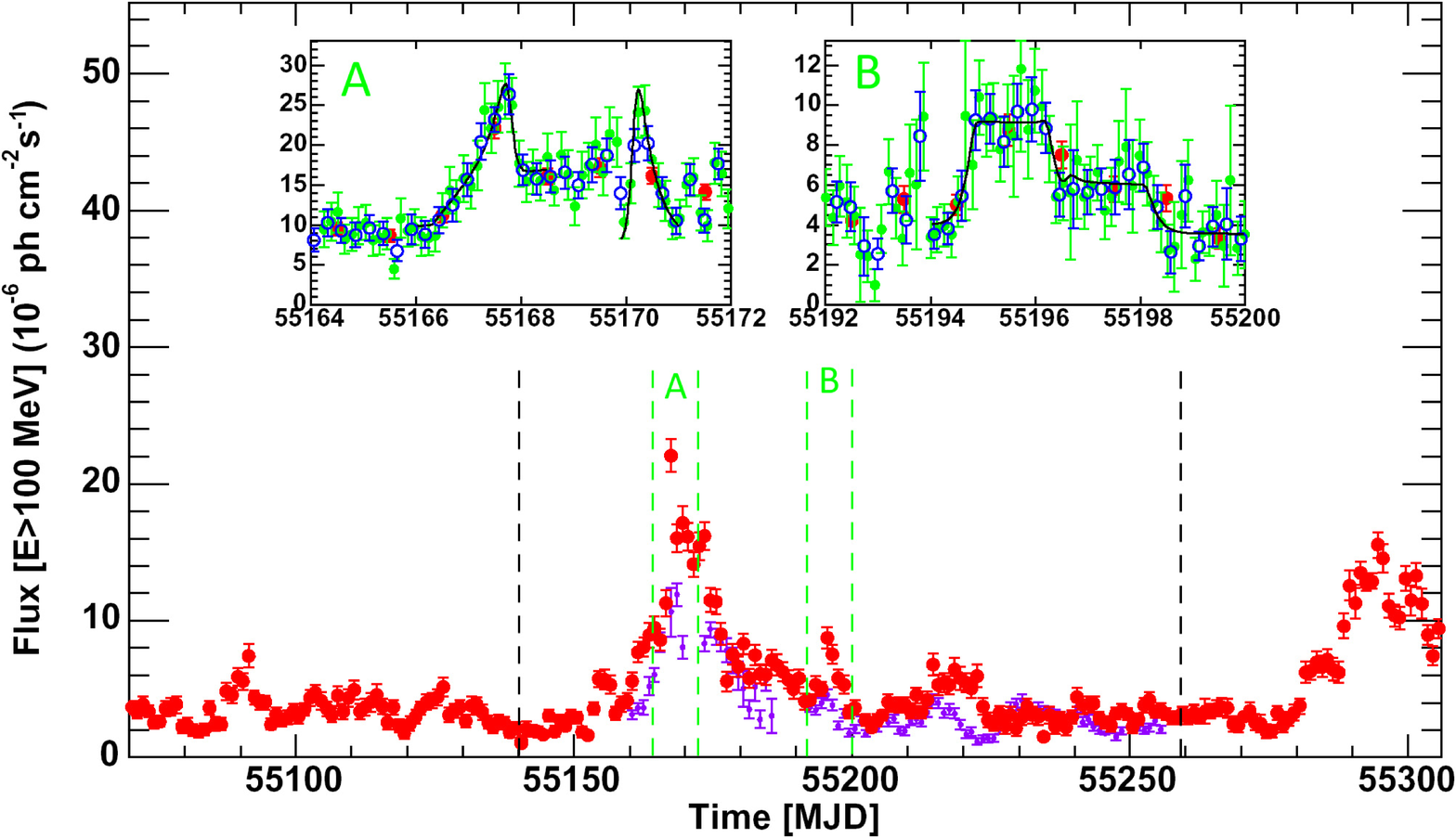}}}
\vskip -0.2cm 
\resizebox{8.6cm}{!}{\rotatebox[]{0}{\includegraphics{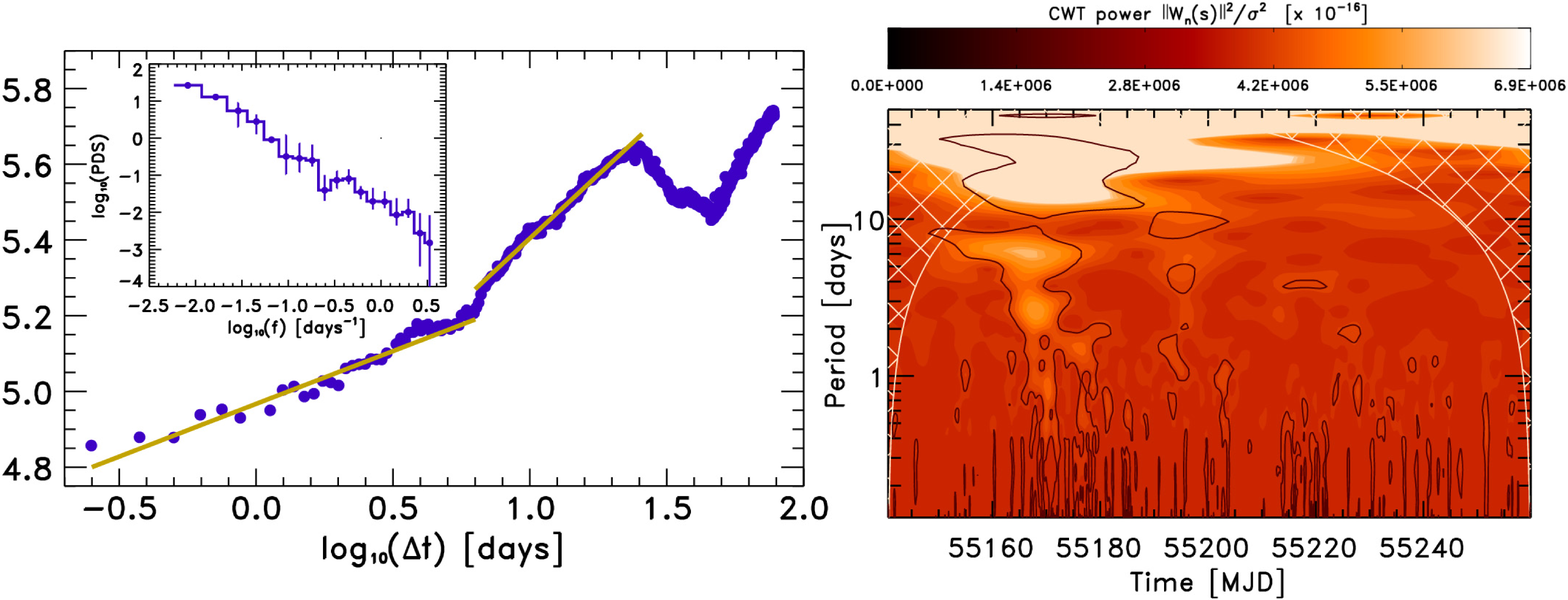}}}
%
\vspace*{-0.3 cm}
\caption{\textit{Top panel:} daily-bin $\gamma$-ray flux light curve of blazar 3C 454.3 (100 MeV - 200 GeV band, red/gray points, main panel) between MJD 55070 and 55307 (2009 Aug. 27 – 2010 Apr. 21). Dashed black
lines mark period over which the SF, PDS and Wavelet analysis are conducted (using improved resolution with 3 hour bins). The light curve of the previous 2008 July–Aug. flare, shifted by 511 days, is also shown (violet/dark-grey tiny points). Insets show blow-ups of the two periods (A and B on the plot) when the largest relative flux increases took place (red/gray filled, blue/dark open, green/light-gray filled data points corresponding to daily, 6-hour, 3-hour bin fluxes, respectively).
\textit{Bottom left panel:} SF of the 3h-bin flux light curve for the period MJD 55140-55259 (2009 Nov. 5 -
2010 Mar. 4, black dashed lines in the top panel) and corresponding PDS (inset). \textit{Bottom right panel:} 2D contour plot of the continuous Morlet wavelet transform power density of the same light curve (thick black contours: 90\% confidence levels of true signal features against white-noise and red/flickering noise background; cross-hatched region: cone of influence dominated by spurious power of time-frequency edges). From \citet{LAT_3C454}.}
  \label{fig:3C454}
\end{center}
\vspace*{-1.0 cm}
\end{figure}
%

The entire $\gamma$-ray is observed every 2 orbits ($\sim 3$ hours) representing a continuous monitor for variability and transients allowing the collection of regular daily/weekly-sampled light curves for dozens of GeV sources. Multiwavelength (MW) observing campaigns are therefore limited only by the source brightness and by the ability to coordinate other telescopes (Fig. \ref{fig:SED-PDS_plane}).

%
\begin{figure*}[th!!!]
\begin{center}
\vspace*{-0.1 cm}
\resizebox{16.0cm}{!}{\rotatebox[]{90}{\includegraphics{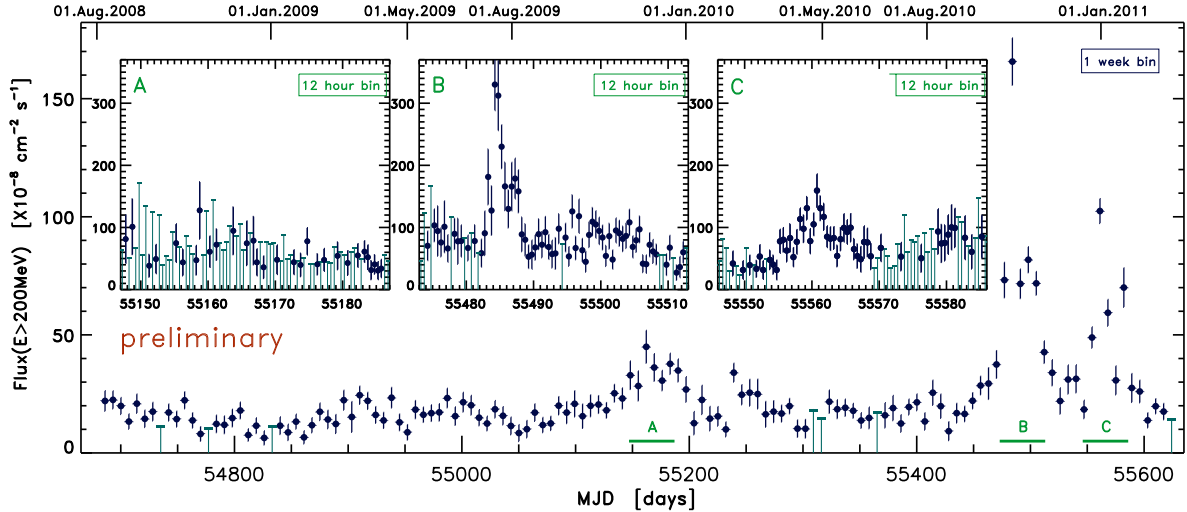}}}
\vspace*{-4.5 cm}
\caption{\textit{Main panel}: 31.5-month (945-day) light curve of the  integral $\gamma$-ray flux above 200 MeV in weekly time bins of blazar PKS 1830-211, from 2008 August 04, to 2011 March 7 (MJD 54682.65 to 55627.65). This is an example of the \fermi LAT capabilities in high-energy temporal variability monitoring.
\textit{Inset panels}: 12-hour bin light curves detailing the period around the mild flare of October 2009 (A interval), detailing the period around the large outburst of October 2010 (B interval) and the secondary double flare of December 2010 and January 2011 (C interval).}
  \label{fig:PKS1830}
\end{center}
\end{figure*}
%
\begin{figure*}[th!!!]
\vspace*{-0.2 cm}
\begin{center}
\hskip -0.2cm 
\resizebox{\hsize}{!}{\rotatebox[]{0}{\includegraphics{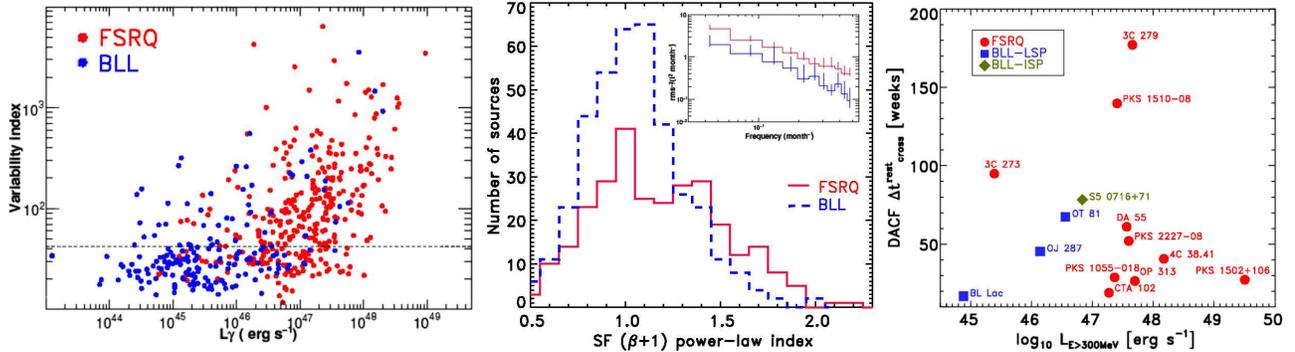}}}
\hskip 0.3cm 
\vspace*{-0.8 cm}
\caption{Results from $\gamma$-ray light curves extracted in fixed 1-month bins over the first 2 years of \fermi LAT all-sky survey for the 886 blazars/AGNs of the clean source sample of the second \fermi LAT Source Catalog, 2FGL. \textit{Left panel}: variability index versus isotropic $\gamma$-ray luminosity (red/grey: FSRQs; blue/black: BL Lac Objects). Dashed line represents the 99\% confidence level for a source to be variable. \textit{Central panel}: distribution of the temporal PDS power-law indexes ($\alpha = \beta + 1$) for the FSRQs (red/continuous line) and BL Lac objects (blue/dashed line) of the sample evaluated in time domain using first order structure
function (SF) analysis (blind power-law index $\beta$ estimation using a maximum lag of 2/3 of the total light curve range). Cumulative power density spectra (PDS) for bright FSRQs (red/top line) and BL Lac objects sub-samples (blue/bottom line) showing similar slopes (inset). From \citet{2LAC}. \textit{Right panel}: scatter plot of the weekly-binned variability auto-correlation (DACF) crossing times in the rest frame of the source (corrected for z and beaming) versus the total apparent isotropic $\gamma$-ray luminosity ($E > 300$ MeV) in the co-moving frame
for 15 bright LAT blazars that are also part of the MOJAVE program. 3C 454.3 is out of the plot range (with $\Delta t^{cross}_{rest} = 254.3$) weeks and $\log_{10}(L_{E}) = 48.1$. Adapted from \citet{2LAC,3monthLBASlightcurves}.
}
\label{fig:3variabilitypanels}
\end{center}
\end{figure*}
%
%
%

\begin{figure*}[t!!!]
\begin{center}
\hskip -0.3cm 
\resizebox{7.2cm}{!}{\rotatebox[]{0}{\includegraphics{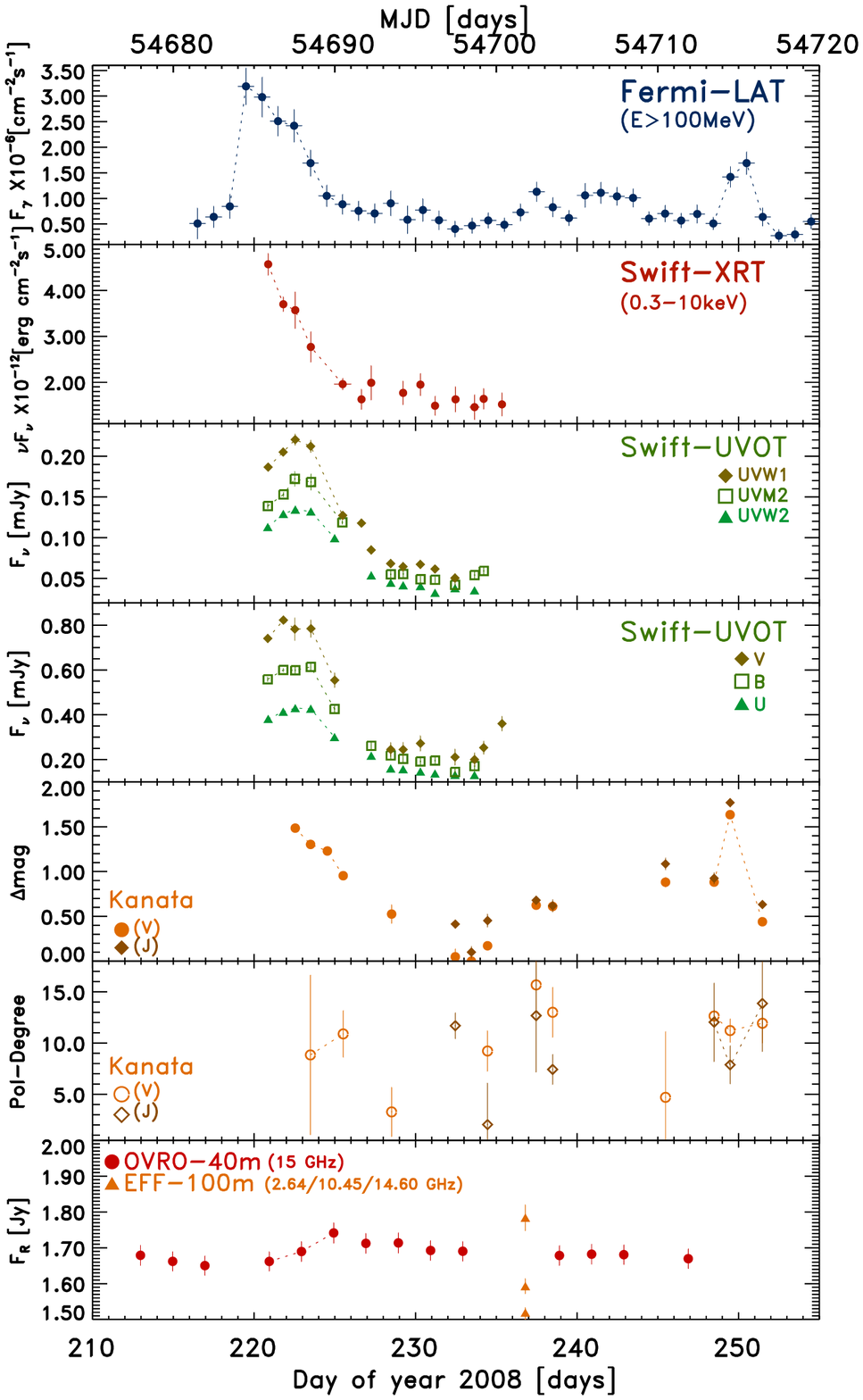}}}
\hskip 0.1cm 
\resizebox{9.2cm}{!}{\rotatebox[]{0}{\includegraphics{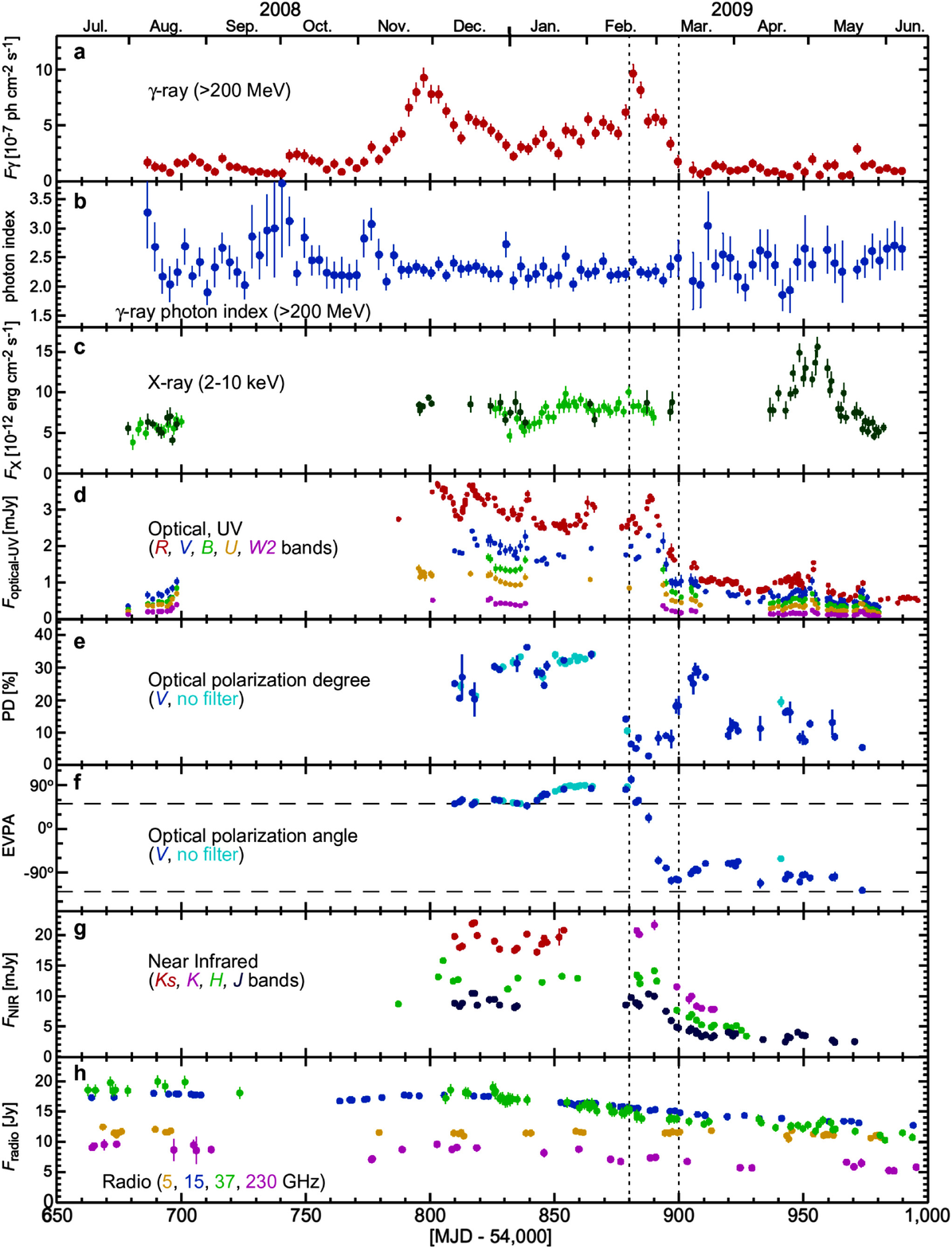}}}
\vspace*{-0.4 cm}
\caption{Radio, mm, optical (flux, polarization), UV, X-ray, $\gamma$-ray fluxes light curves obtained by \fermi coordinated MW campaigns (PKS 1502+106, left, 3C 279, right). Adapted from \citet[]{LAT_PKS1502,LAT_3C279_Nature}.}
  \label{fig:PKS1502e3C279}
\end{center}
\end{figure*}
%

%
\begin{figure}[t!!!]
\begin{center}
\hspace*{-0.5cm} 
\resizebox{9cm}{!}{\rotatebox[]{0}{\includegraphics{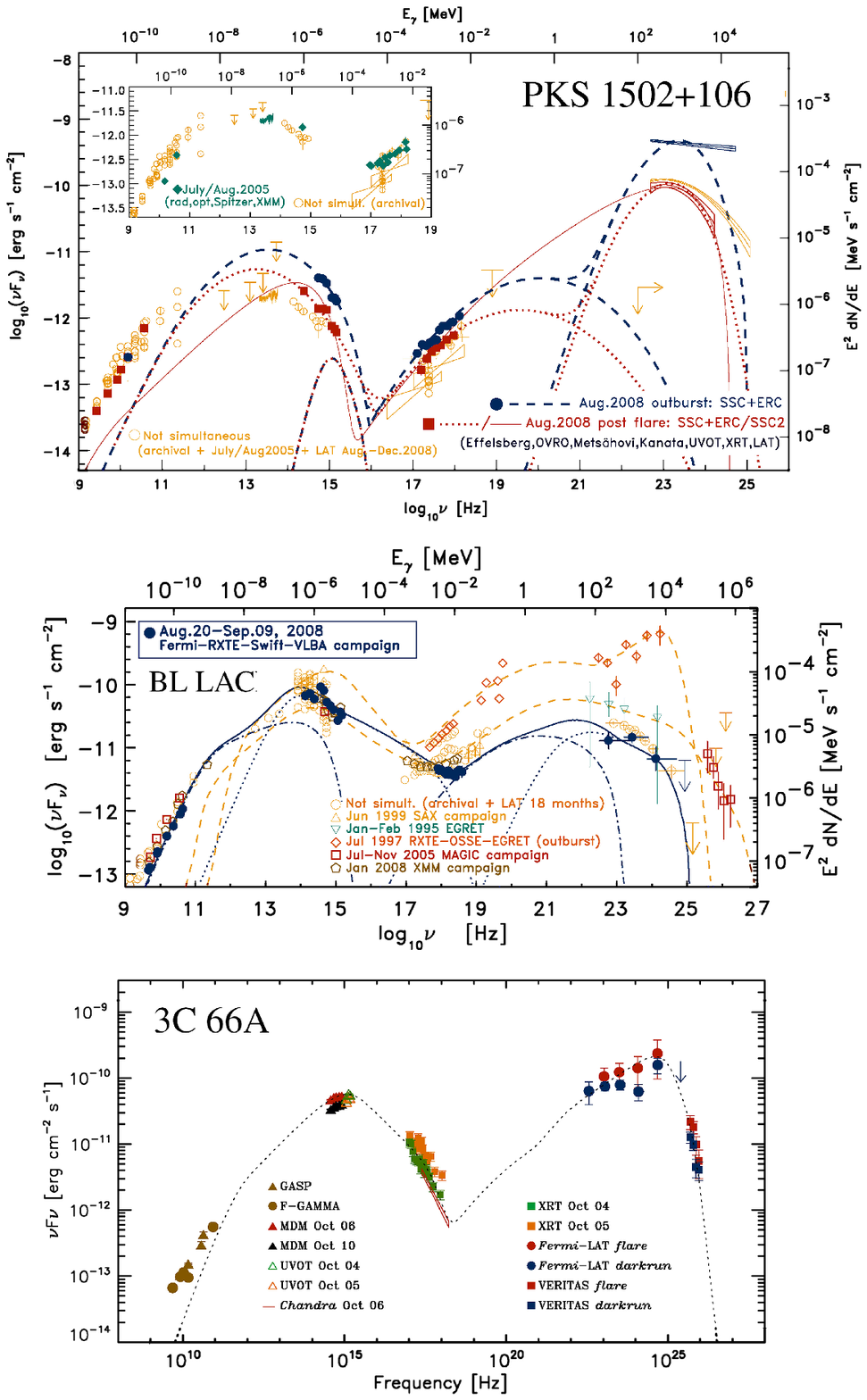}}}
\vspace*{-0.1 cm}
\caption{Multi-epoch spectral energy distributions of blazars (PKS 1502+106, BL Lac, 3C 66A) with multifrequency data
obtained with coordinated and simultaneous MeV-GeV gamma-ray (by \fermi LAT) and radio, mm, optical UV, X-ray and TeV (bottom panel only) observations. Adapted from \citet[]{LAT_PKS1502,LAT_BLLac,LAT_3C66A}.}
  \label{fig:3SED}
\end{center}
\end{figure}
%

Irregular and aperiodic variability is found in blazars at all the timescales and at all the energies (MW variability). EGRET already showed that blazars have a high-energy component in their spectral energy distributions (SED) and are the largest class of variable $\gamma$-ray sources, although it was limited by statistics, while \fermi is observing low $\gamma$-ray brightness states too (example: BL Lac in Fig. \ref{fig:3SED}). Of the studied \fermi LAT blazars, 2/3 are variable and high states are less than 1/4 of the total light curve range \citep{3monthLBASlightcurves}. Flat Spectrum Radio Quasars (FSRQs) and low energy peaked BL Lac objects (BL Lacs) show the largest relative variance (Fig. \ref{fig:3variabilitypanels}), while high-energy peaked BL Lacs display a lower variable but persistent emission. Sources like PKS 1510-08, PKS 1502+106 (Fig. \ref{fig:PKS1502e3C279} and \ref{fig:3SED}), 3C 454.3 (Fig. \ref{fig:3C454}), 3C 279 (Fig. \ref{fig:PKS1502e3C279}), PKS 1830-211 (Fig. \ref{fig:PKS1830}), 4C 21.35, 4C 38.41 (all FSRQs) and AO 0235+164, 3C 66A (both BL Lacs) are among the brightest, most variable, and isotropically luminous
blazars seen by \fermi. For bright flares even intra-day light curves have been extracted as can be seen, for example, in Fig. \ref{fig:3C454} and \ref{fig:PKS1830}.

Discrete autocorrelation function (DACF) and structure function (SF, Fig. \ref{fig:3variabilitypanels} and \ref{fig:3C454}) analysis showed different patterns, autocorrelation times and power-law PDS indices ($1/f^{\alpha}$ fluctuations, where $f=1/t$) implying different variability modes for each source (more flicker, $\alpha \simeq 1$ or more Brownian, $\alpha \geq 2$, dominated). 3C 454.3 is a fully Brownian $\gamma$-ray source while other powerful blazars have values half-way between the two modes. Average power spectral density analysis (PDS) in frequency, $f$, domain over the blazar subclasses points out $\alpha$ slopes from 1.3 to 1.6 \citep{3monthLBASlightcurves,2LAC}. No evidence for persistent characteristic $\gamma$-ray timescale(s) is found. Flare profiles are mostly symmetric (in part because of superimposing flares and large bin smoothing). The fractional variability during outburst appears similar to its longer term mean in the few objects studied in detail.

%
%
\section{Multi-waveband blazar variability through observing campaign led by \fermi}    
%
%

\textit{Fermi}-driven MW observing campaigns are shedding light on the PDS-SED plane (i.e. timescale-energy parameter space).
Broad-band MW studies are addressed mostly to cross-correlation and time-lag analysis, to time-resolved SED modeling, to the search for orphan flares and spectral hysteresis, to the analysis of $\gamma$-ray vs synchrotron amplitude ratios and emission peaks and to the study of the radio-$\gamma$-ray connection and  source populations. Simultaneous MW observations are crucial also for the identification of newly discovered $\gamma$-ray sources.

\par In order to define and better constrain physical parameters, processes and emission components, to clarify the role of the central engine, jets and their interplay, the jet composition and structure in AGNs and blazars is necessary to collect more different and longer sequences of MW observations. Some clues are already emerging.\\
1) The knowledge of redshifts is crucial but $\sim 50\%$ of BL Lacs have still unknown $z$. 2) Simple single-zone synchrotron self Compton (SSC) descriptions are vanishing. 3) Internal shock scenario, eventually with composite particle energy distributions, works well. 4) Cross-correlation analysis with optical polarization provides important clues on jet physics. 5) The location of emission site can be both inside and outside the broad line region (BLR). 6) Magnetic fields are complex but can be highly ordered and jet-aligned during $\gamma$-ray flares. 7) In some cases bright $\gamma$-ray flares seem to occur after ejections of superluminal radio knots.
Two examples of composite MW light curves obtained from \fermi blazar campaigns (PKS 1502+106 and 3C 279) are reported in Fig. \ref{fig:PKS1502e3C279}, \cite{LAT_PKS1502,LAT_3C279_Nature}.

In conclusion \fermi LAT, in the frame of blazars and other AGNs science, is demonstrating very good capabilities in the field of $\gamma$-ray variability analysis and radio-gamma-ray connection and is showing an optimal synergy with the \textit{Swift} mission.

\begin{acknowledgments}
\footnotesize{The \fermi LAT Collaboration acknowledges support from a number of agencies and institutes for both development and the operation of the LAT as well as scientific data analysis. These include NASA and DOE in the United States, CEA/Irfu and IN2P3/CNRS in France, ASI and INFN in Italy, MEXT, KEK, and JAXA in Japan, and the K.~A.~Wallenberg Foundation, the Swedish Research Council and the National Space Board in Sweden. Additional support from INAF in Italy and CNES in France for science analysis during the operations phase is also gratefully acknowledged.}
\end{acknowledgments}

\bigskip 

\end{document}